\documentclass[]{spie}  %>>> use for US letter paper
\usepackage[]{graphicx}

\usepackage{cite}
\usepackage[tight,footnotesize]{subfigure}
\usepackage[cmex10]{amsmath}
\usepackage{amsfonts}
\usepackage{algorithm}
\usepackage{algpseudocode}
\usepackage{algorithmicx}

\DeclareMathOperator*{\argmin}{\arg\!\min}
\DeclareMathOperator*{\sfx}{\text{\textit{sfx}}}

\title{Free-body Gesture Tracking and Augmented Reality Improvisation for Floor and Aerial Dance} 

\author{Tammuz Dubnov\supit{a} and Cheng-i Wang\supit{b}
\skiplinehalf
\supit{a}UC San Diego, CA, USA\\ 
\supit{b}Calit2, UCSD, La Jolla, CA, USA
}

\authorinfo{Further author information: (Send correspondence to T.Dubnov)\\E-mail: tdubnov@eng.ucsd.edu, Telephone: 1 619 895 9950}
\begin{document} 
  \maketitle

%%%%%%%%%%%%%%%%%%%%%%%%%%%%%%%%%%%%%%%%%%%%%%%%%%%%%%%%%%%%% 
\begin{abstract}
This paper describes an updated interactive performance system for floor and Aerial Dance that controls visual and sonic aspects of the presentation via a depth sensing camera (MS Kinect). In order to detect, measure and track free movement in space, 3 degree of freedom (3-DOF) tracking in space (on the ground and in the air) is performed using IR markers with a method for multi target tracking capabilities added and described in detail. An improved gesture tracking and recognition system, called Action Graph (AG), is described in the paper.  Action Graph uses an efficient incremental construction from a single long sequence of movement features and automatically captures repeated sub-segments in the movement from start to finish with no manual interaction needed with other advanced capabilities discussed as well. By using the new model for the gesture we can unify an entire choreography piece by dynamically tracking and recognizing gestures and sub-portions of the piece. This gives the performer the freedom to improvise based on a set of recorded gestures/portions of the choreography and have the system dynamically respond in relation to the performer within a set of related rehearsed actions, an ability that has not been seen in any other system to date.
\end{abstract}

%>>>> Include a list of keywords after the abstract


%%%%%%%%%%%%%%%%%%%%%%%%%%%%%%%%%%%%%%%%%%%%%%%%%%%%%%%%%%%%%
\section*{INTRODUCTION}
\label{sec:intro}  % \label{} allows reference to this section

The paper describes technology for an interactive multimedia composition for dancer or aerial performer and a computer that utilizes audiovisual materials that are controlled by her/his movements. These movements are detected using a RGBD camera that tracks infra-red (IR) markers that are placed on the performer`s body \cite{Zhang2012}. The 3D information about the position of the markers in space are input into a tracking and recognition system called Action Graph that uses models based on similar gestures that were recorded at a training phase. During the performance the dancer's movements influence the audiovisual and musical elements by controlling the speed of the playback of the audiovisual materials, as well as generating additional graphics and transforming pre-designed video and sound playback materials in response to the dancers movements.
In a previous paper \cite{Dubnov2014}  we described various strategies of mapping gestures and movements to graphics and discussed the considerations behind composition in an augmented reality setting \cite{DTZ}. One of the limitations of the previous system was the inability to effectively track multiple markers simultaneously, leading to major issues and limitations in terms of on the scope of gestures and freedom of movements for the performers wearing the markers. In this paper we describe an expanded version that is able to dynamically track multiple IR markers and deals with problems of occlusion (marker disappearance) and reappearance, as well as crossing trajectories and other disturbances that are commonly experienced in a live movement performance.  

One of the difficulties of integrating motion detection and recognition into a dance and aerials performance is the large variability of gestures and freedom in choice of the movements by the dancers and the choreography choices. In our previous system we used a left-to-right HMM \cite{Bevilacqua2010} trained on separate segments of the performance or individual movement sequences. This was fairly limiting as the system could detect individual segments but would need manual interactions to indicate when to attempt detection within the performance. It is desirable for the system to be able to identify typical subsequences in a single long recording of a dance, without breaking it up into individual scenes. In this paper we propose an alternative approach to movement modeling that we call Action Graph that allow an efficient incremental construction from a single long sequence of movement features and that captures repeated sub-segments in the movement from start to finish with no manual intervention. The model, based on Variable Markov Oracle (VMO) \cite{Wang2014} has additional advantages, such as direct mapping of the performance timeline, and its associated graphics and video element, to movement primitives (gestural subsequences) that are found in that particular dance.

Finally, we  outline the steps and usage of this software in production, from choreographing, rehearsing and finally performance phases. By using the new model for the gesture we can unify an entire choreography piece, and potentially an entire show, with the model dynamically tracking and recognizing gestures and sub portions of the piece. This can give the performer the freedom to improvise from a set of recorded gestures/portions of the choreography and have the system dynamically change and control/intelligently improvise in relation to the performer within the framework of related rehearsed choreography, an ability that has not been seen, to the best of our knowledge, in any other system to date.

%An upcoming performance is schedule to occur at UCSD through the IDEAS program in Calit2 \cite{Dubnov2014} that will allow local dancers and aerialists to interact with the system and make their own augmented reality choreographies. Results and videos of rehearsals with the system, which we named Zuzor, can be found at http://shlomodubnov.wikidot.com/kinect-aerial-dance or https://www.facebook.com/ZuzorKinect.

%%%%%%%%%%%%%%%%%%%%%%%%%%%%%%%%%%%%%%%%%%%%%%%%%%%%%%%%%%%%%

\section*{MULTIPLE IR MARKERS}

Similar to the previous paper, our tracking algorithm uses two pieces of information from the depth camera: an IR image and a depth image. Our method for extracting the marker coordinate uses a right-handed coordinate system with z pointing directly out from the camera, y pointing out of the top of the camera, and x pointing to the right of the camera, when facing it. For specifics of marker coordinate extraction from the IR and depth image please refer to the previous paper\cite{Dubnov2014}. The main limitation of our previous tracking algorithm is that it tracked and sorted the blobs (IR markers) based on their depth (z coordinate), effectively meaning that if two markers pass one another in the z axis then the tracking algorithm would confuse the two markers with one another and switch their labels. This clearly put a great limitation on the capabilities of that system that handicapped the choreographers, graphic designers and performers involved. This limitation is overcome by tracking of the markers using the so called Kuhn-Munkres algorithm, with further versatility supported by additional logic in cases of death (no longer visible to the depth camera) and rebirth (as they come into view) as well as crossing trajectories and other disturbances that may occur.

%As described in the  following subsections, the Hungarian algorithm is used to best match blobs from frame to frame and the added logic to intelligently handle as markers disappear (such as cases of marker occlusion such as being covered by other props or performers on stage or going off stage) and reappear (new markers coming entering the space visible to the RGB-D camera).

\subsection*{Kuhn-Munkres Algorithm}
%NOTE I WILL AM TALKING ABOUT WHAT THE HUNGARIAN ALGORITHM DOES AND HOW IT CAN BE IMPLEMENTED IN OUR CASE RATHER THAN HOW THE ALGORITHM WORKS IN DETAIL!! NEED TO MAKE SURE THAT THAT IS OK AND CAN ADD IT LATER IF WE NEED MORE PAGES//

The Kuhn-Munkres (KM) algorithm (also known as the Hungarian algorithm)  is a combinatorial optimization algorithm that solves the assignment problem in polynomial time. Since it is well documented, we will not go into much details that appear in the original paper \cite{Munkres1957}, but will instead highlight how it can be implemented to solve the multi-target issue at hand.

The KM algorithm can solve problems of minimization between a bipartite graph by considering non-negative edge costs and returning the graph assignment with the minimum accumulative edge costs. In our situation we would consider the most recent frame, which we will refer to as frame $n$, and the current preprocessed frame, which we will refer to as frame $n+1$, with their respective IR markers as the two partitions of the graph. The edge costs are defined as the 3D distance between the coordinates of every blob from frame $n$ and every blob from frame $n+1$. We will be using the trivial 3D distance calculation between blob $i$ and blob $j$ defined as follows
$$ \Delta_{ij} = \sqrt{(x_i-x_j)^2+(y_i-y_j)^2+(z_i-z_j)^2}.$$

\begin{figure}
\centering
\includegraphics[scale=.6]{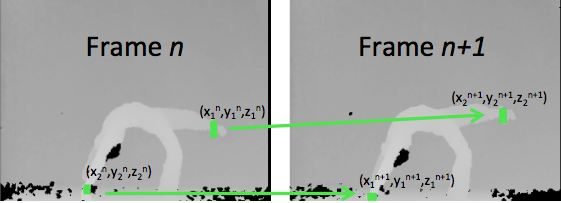}
\caption{Two blobs visible in two consecutive frames, notice that the indexing of the two blobs is switched between Frame $n$ and Frame $n+1$.}
\label{fig:ex1}
\end{figure}

Assuming that all markers on stage are worn by performers and remain visible to the RGB-D camera at all times, then the KM algorithm should effectively match the blob at time $n$ to it's new position in time $n+1$ in the new frame. Since the sampling is done at a rate of 30 frames per second, it us unlikely that two non overlapping moving blobs would move fast enough or cross each other to new positions that would be closer to the other's previous location. Such an example can be seen in Figure \ref{fig:ex1}.

The figure shows frame $n$ with the top marker indexed as "1" and the bottom as "2", and frame $n+1$ with the top marker indexed as "2" and the bottom as "1". The KM algorithm matches them as illustrated since the distance between each blob's position at time $n$ versus $n+1$ is minimal compared to the distance between the different markers and their other. Thus it would overcome the index confusion and correctly sort them.

\subsection*{Death and Rebirth Logic}
\begin{figure}
\centering
\includegraphics[scale=.6]{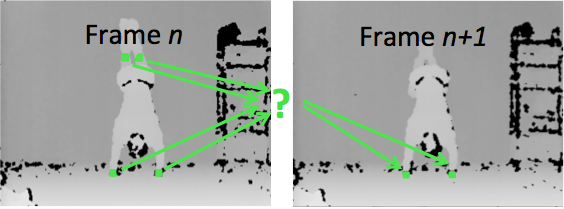}
\caption{Of the four blobs visible in Frame $n$, two disappear in Frame $n+1$ and extra logic is needed to intelligently update the indexes accordingly.}
\label{fig:ex2}
\end{figure}

\begin{figure}
\centering
\includegraphics[scale=.6]{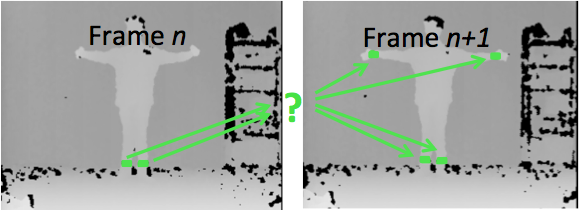}
\caption{In Frame $n$ two blobs are visible and two are covered by the angle of the arms thus two blobs are born in Frame $n+1$ and the previous tags of the initially visible blobs must be conserved.}
\label{fig:ex3}
\end{figure}

As markers constantly and continually come in and out of consideration due to occlusion, noise or going on/off line of sight, additional logic is required.
%A simple example would be a marker being shortly obstructed due to being briefly covered, and the blob associate with the marker change every time the marker is shortly unseen. We have to cases to consider in developing a scheme that would intelligently/logically track the blobs as markers come in and out of view.
We have the death case demonstrated in Figure \ref{fig:ex2} in which two markers disappear and an unsupervised/unspecified tracker using KM could accurately track which markers are continued but have the index of the blobs mixed up. The second case is rebirth in which new markers are visible in the new frame and the tracker needs to be able to track the old markers accurately as well as initiate the marker tracking for the newly visible markers and correctly index the blobs. This can be seen in Figure \ref{fig:ex3}.

The process of keeping track of death and rebirth is outlined as follows:
A number of maximum number blobs planned to be on stage in provided. These are stored in a dictionary with their coordinates (initially defined to be all {\it{inf}}), and a list of labels. We keep a states list, with "alive" for blobs actively seen and "dead" for blobs not currently seen by the sensor. We then have three fundamental action possibilities for every blob, "death", "birth" and "living", which are maintained as we progress from frame to frame, and an extra optional "kill" action defined as well.

At each step we run the KM algorithm between all the past locations of "alive" blobs and the incoming marker coordinates. If additional incoming markers exist that are not assigned to previously "alive" blobs, then the KM algorithm is run twice: first it is applied between a list of "dead" blobs and the remaining unused incoming marker coordinates, and secondly it is used to match between the incoming markers and a combined list of "alive" and "dead" blobs. We then consider the two costs and chose the least one, possibly weighting the first cost more heavily.
The intuition for weighting the first cost more heavily is because it seems be more likely to have a blob remain alive between every frame rather rather than die at the same frame that a new one is born.
According to the winning cost, if the blob was alive and is again found in the new frame then a "living" action is executed updating that blob's coordinate in the dictionary, with the blob remaining in the "alive" list. If the blob was alive and isn't found in the new frame then a "death" action is executed in which the coordinate isn't updated and the blob is moved form the "alive" list to the "dead" list. If the blob was dead and is now alive (found to be in the new frame) then a "birth" action is executed updating the blob's coordinate in the dictionary and moving it from the "dead" list to the "alive". If the blob was dead and is not found in the new frame then no action is taken on the blob.
The additional "kill" action is only used if there are more incoming marker coordinate than the maximum number of blobs specified, in which case the KM algorithm would be run effectively matching all the blobs to their optimal next frame coordinate, and the extra incoming marker coordinates would be assumed to be noise and would be "killed" and remove from the list.

%%%%%%%%%%%%%%%%%%%%%%%%%%%%%%%%%%%%%%%%%%%%%%%%%%%%%%%%%%%%%

\section*{ACTION GRAPH UTILIZING VMO}
After the blobs are detected from the multiple IR markers, the sequence of blob frames effectively forms a multivariate time series. To track the gestures embedded in the multivariate time series, the \emph{Variable Markov Oracle} (VMO) \cite{Wang2014} is used as the core data structure enabling the online tracking algorithm. The gestures in the task are defined as repeated sub clips of the original multivariate time series. In previous work, the gesture tracking is done via the left-to-right HMM proposed in \cite{Bevilacqua2010}. The motivation for replacing the left-to-right HMM with VMO is to avoid the need of manually deciding the timing to start gesture tracking. Other works on Kinect gesture tracking recognition mostly rely on the skeletal information \cite{LaViola2013,Zhao2013} hence are different from our approach where the raw IR images are used. 

VMO is a suffix tree data structure for multivariate time series originally devised for query-guided audio content generation \cite{Wang2014} and multimedia query-matching \cite{Wang2014a}. The query related tasks were suitable for VMO since VMO is capable of finding embedded clusters of observations along the multivariate time series and thus enables a Viterbi-like dynamic programming query-matching algorithm to be devised. In this work, the offline query-matching algorithm proposed in \cite{Wang2014,Wang2014a} is adapted to an online version for the gesture tracking and recognition task defined in this paper. The properties of VMO are briefly described in this section but for details about the construction and model selection algorithms of VMO please refer to \cite{Dubnov2011,Wang2014}. 

\begin{figure}
\centering
\includegraphics[width=0.6\columnwidth]{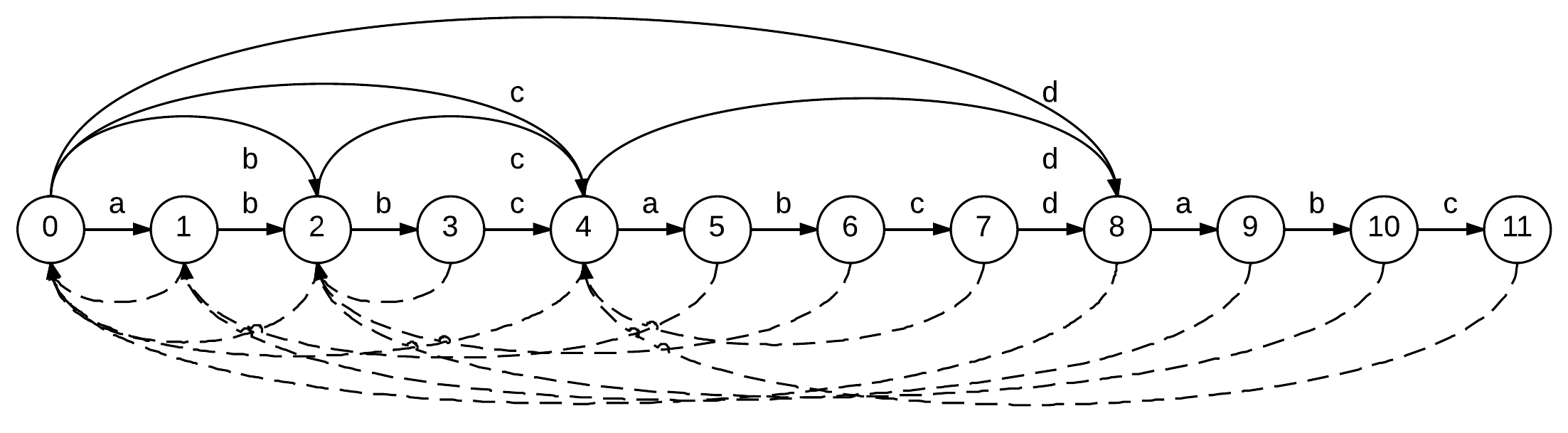}
\caption{(Top) A \emph{VMO} structure with symbolized signal $\{a,b,b,c,a,b,c,d,a,b,c\}$, upper (normal) arrows represent forward links with symbols for each frame and lower (dashed) are suffix links.}
\label{fig:vmo}
\end{figure}

An example of the VMO data structure is depicted in Figure \ref{fig:vmo}. VMO clusters a multivariate time series $O$, sampled at time $t$, into a symbolic sequence $Q=q_1,q_2,\dots,q_t,\dots,q_T$, with $T$ states and with frame $O[t]$ represented by a symbol $q_t$. The symbols are formed by tracking suffix links along the states in an oracle structure. A VMO carries two kinds of links; one is forward link and the other is suffix link. Suffix link is a backward pointer that links state $t$ to $k,t>k$, without a label and is denoted by $\sfx[t]=k$. 
\begin{align*}
\sfx[t]=k\iff & \text{the longest repeated suffix of}\\ 
&\{q_1,q_2,\dots,q_t\}\text{ is recognized in }k.
\end{align*}
Suffix links are used to find repeated patterns in $Q$. 

Two types of forward links are presented in the structure; the first is an internal forward link which is a pointer from state $t-1$ to $t$ labeled by the symbol $q_t$, denoted by $\delta(t-1,q_t)=t$. The other forward link is an external forward link which is a pointer from state $t$ to $t+k$ labeled by $q_{t+k}$ with $k>1$. An external forward link $\delta(t, q_{t+k}) = t+k$ is created when 
$$ q_{t+1}\ne q_{t+k}\wedge q_{t}= q_{t+k-1}\wedge \delta(t, q_{t+k})=\emptyset.$$
In other words, an external forward link is created when the newly added internal forward link is unseen for previous occurrence of $q_t$. The function of the forward links is to provide an efficient way to retrieve any of the factors of $Q$, starting from the beginning of $Q$ and following a unique path formed by forward links.

In order to fulfill the online requirement for the gesture tracking in this work, the query-matching algorithm proposed in \cite{Wang2014} is revised. The online algorithms are provided in Alg. \ref{alg:tracking_init} and Alg. \ref{alg:tracking_online}. Basically, the online gesture tracking is done by a frame-by-frame basis instead of a backward pass at the end of the query time series for the offline version. In Alg. \ref{alg:tracking_init}, $K$ possible gesture candidates are initialized according to the number of clusters discovered by VMO and the first index for each gesture candidates is located according to the distance between the first observation from the input stream to every frame in each cluster. Each cluster of frames, $b_k$, is a list of pointers to the frames connected by one unique suffix path found by VMO. For notation purposes, a list of the cluster lists is notated as $B$. A path vector $M$ and cost vector $C$ are returned from Alg. \ref{alg:tracking_init} and passed to the online tracking algorithm. In Alg. \ref{alg:tracking_online}, for each incoming observation, the algorithm is run once to update $M$ and $C$ for each gesture candidate, then $M_{\argmin{C}}$ is returned as the found index in target time series $P$ indicating the best match between the incoming time series $R$ and stored time series $O$. The VMO and its accompany gesture tracking algorithm together form the Action Graph system proposed in \cite{Dubnov2014}. 

\begin{algorithm}
\begin{algorithmic}[1]
\Require{Multiple Gestures stored \emph{VMO}, $P$, includes $Q$ the cluster label sequence and $O$ the signal and first frame of incoming time series $R_1$}
\State Get the number of clusters, $K\gets|B|$
\State Initialize cost vector $C\in\mathbb{R}^K$ and path vector $M\in\mathbb{R}^{K}$ with all zeros.
\For{$k=1:K$} 
\State $D\gets$ all distances between $R_1$ and $O[b_k]$
\State $M_k=b_k[\argmin(D)]$
\State $C_k = \min(D)$
\EndFor
\State \Return $M_{\argmin{C}}$, $M$, $C$ 
\end{algorithmic}
\caption{Gesture Tracking - Initialization}\label{alg:tracking_init}
\end{algorithm}

\begin{algorithm}
\begin{algorithmic}[1]
\Require{Multiple Gestures stored \emph{VMO}, $P$, path vector $M$, cost vector $C$ from Alg \ref{alg:tracking_init} and incoming time series $R_t$ at time $t$.}
\For{$k=1:K$}
\State Gather the cluster labels for forward links 
\Statex \hskip\algorithmicindent from $M_{k}$ and $M_{k}$ itself.
\Statex \hskip\algorithmicindent $\eta\gets Q[\delta(M_{k}); M_{k}]$
\State Gather all states from possible clusters, 
\Statex \hskip\algorithmicindent $b'\gets B[\eta]$ 
\State $D\gets$ all distances between $R_t$ and $O[b']$
\State $M_{k}=b'[\argmin(D)]$
\State $C_k \mathrel{{+}{=}} \min(D)$
\EndFor
\State \Return $M_{\argmin{C}}$, $M$, $C$ 
\end{algorithmic}
\caption{Gesture Tracking - On-line Tracking}\label{alg:tracking_online}
\end{algorithm}

\begin{figure}
\centering
\includegraphics[width=0.8\columnwidth]{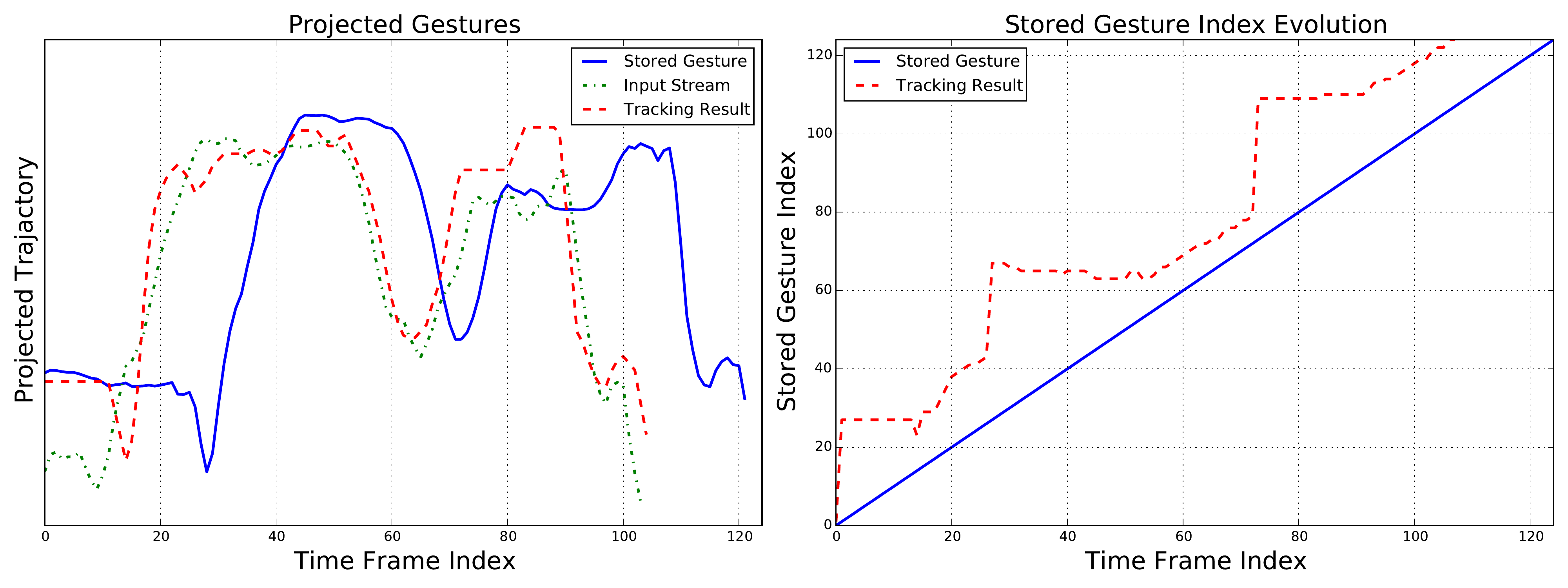}
\caption{(Left) Projected trajectory of the stored gesture (blue), input stream (green) and tracking result (red). (Right) Time indices evolution of the tracking result (red) compared to original gesture (blue).}
\label{fig:tracking_ex}
\end{figure}

Given the online gesture algorithms based on VMO, it is possible to track where in the stored multivariate time series that the newly input observation is closest to. A visualization of such tracking is depicted in Figure \ref{fig:tracking_ex}. In Figure \ref{fig:tracking_ex}, the left sub figure shows the result of the tracking (red dashed line) based on a stored gesture (blue line) given unknown input observations (green dotted line). The right subfigure of Figure \ref{fig:tracking_ex} shows how the tracking result (red dashed line) in relations to its original sequence (blue line) in terms of the matching in time (as of frame indices). The original sequence shown in the figure is part of a longer sequence where multiple gestures sampled from real world gesture data were concatenated into one longer sequence to resemble how it would be used in real world performance environment. This long time series is then projected onto its first principal component. The part shown (blue line) in the figure is one identified gesture (sub clip) found via VMO with suffix links, which might be different from the segmentation by how the time series is concatenated. From the example shown in Figure \ref{fig:tracking_ex}, it could be observed that the on-line tracking algorithm is able to find the correct segment in the stored gesture matching the input observations, though due to visualization purposes, the rest of the stored gesture  (blue line), where the other gestures are located in the original time series, in Figure \ref{fig:tracking_ex} are omitted. Also it could be noticed that from time index evolution in the right subfigure of Figure \ref{fig:tracking_ex}, the tracking is consistent with the temporal relations of the stored gesture (the red dashed line mostly follows the diagonal with time stretch or compression movements indicated by nearly vertical or horizontal lines). 

%%%%%%%%%%%%%%%%%%%%%%%%%%%%%%%%%%%%%%%%%%%%%%%%%%%%%%%%%%%%%

\section*{PRODUCTION METHODS AND PRACTICES}

\begin{figure}
\centering
\includegraphics[width=0.6\columnwidth]{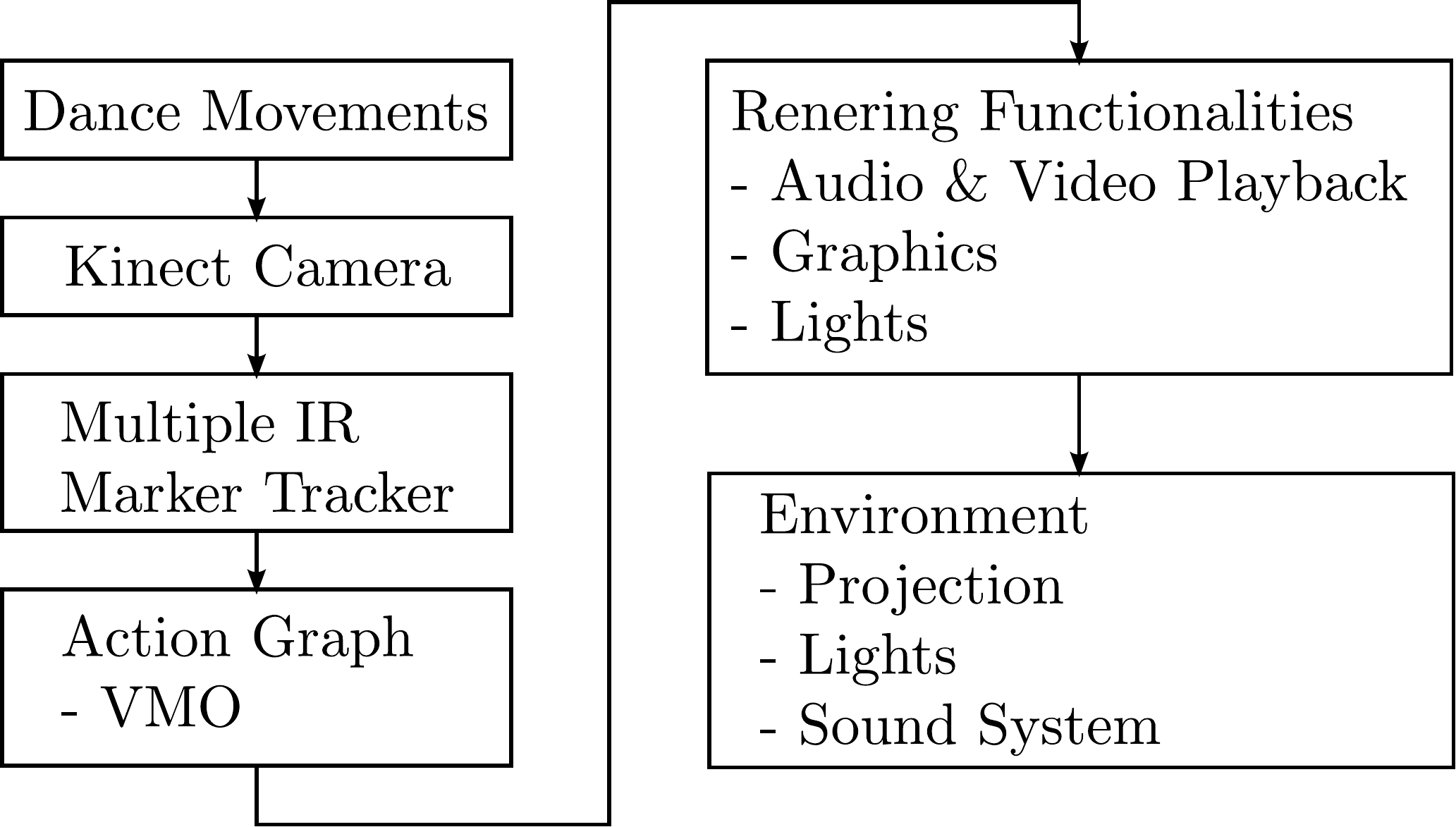}
\caption{Diagram of the proposed system.}
\label{fig:system}
\end{figure}

The aforementioned multiple IR marker tracking techniques and Action Graph are combined into a system depicted in Figure \ref{fig:system}. The Action Graph is used as the mapping interface between the input dance movements and graphics/interaction/effects rendering during a performance. 

For the system proposed in this work, we consider a scenario where the rehearsal or practice of the dancer is recorded and stored with Action Graph as VMO, then the gestures performed by the dancer are defined as the sub clips identified by VMO via suffix links, lastly during the performance the input dance movements are tracked and matched to the gestures stored in VMO in Action Graph. For graphics/effects/interaction rendering, the identified gestures in VMO could be indexed and considered as objects then mapped to the desired rendering mechanisms. Two kinds of mappings are possible with these identified gestures, the first is categorical mapping of different gestures to different rendering functionalities, such mapping could be useful when using different gestures to trigger the on/off of different effects is desired. The second is temporal mapping between one gesture to another time series data, so that the evolvement of an identified gesture could be used to \emph{scrub} another time series, such as video, audio or rendering function parameters that varies in time.

This can be discussed further in relation to various choreography development styles. Since Action Graph is versatile enough to handle multiple sequences individually yet simultaneously or alternatively one length sequence, the choreographer has several options of development with the system. If the choreographer prefers to fully develop shorter sequences for the piece and only towards the end piece them together then he or she can fully develop those section with Action Graph and the accompanying graphics develop and synced to the movement sequence model for every sequence then for the final entire piece the system can automatically stitch all the sequences together and actively identify/recognize which sequence is which and display the corresponding audiovisual content. If alternatively the choreographer prefers to choreograph from beginning to end then Action Graph can support that as well and the choreographer can work with it every step of the way. 

Other options of development could be pre-physical performer choreography, such as the choreographer firstly developing the audiovisual timeline with the song used. After that is finished they could go on and choreograph the dance/physical portion and match the VMO model generated with the audiovisual timeline. This could give additional freedom to alternate between dominantly dancer controlled visuals to dominantly music or timeline controlled visuals. Increased emphasis would be liked to placed on the many different ways in which this system can be used in hand with the choreographer in a non limiting and in fact potentially inspiring fashion for the choreographer and performers alike. 

%%%%%%%%%%%%%%%%%%%%%%%%%%%%%%%%%%%%%%%%%%%%%%%%%%%%%%%%%%%%%

\section*{CONCLUSION}

% a paragraph concludes on multiple IR markers algorithm first maybe?
Steps towards moving away from the current use of audiovisuals with floor and aerial dance in which either the visuals must be general enough to allow for human error while performing or the performer is expected to be incredibly well rehearsed to a degree where they are expected to perform with impossible precision to truly complement and work with the visuals employed. This paper discussed an expansion in tracking capability, both in terms of the number of markers/blobs used and the intelligent indexing logic used to identify them throughout time. This enables further freedom for the choreographer and is more forgiving to human error/variability during live performances by enabling greater accuracy in tracking the performer and finer details within the dance piece. 

We also proposed and showcased how the VMO could be used in Action Graph. The newly proposed Action Graph is capable of tracking and recognizing incoming gestures in an unsupervised way, and enabling the mapping between input gestures to desired rendering functionalities. The unsupervised fashion of tracking and recognition allows for a performance setting where the choreography is allowed to have the form of structured improvisation or variations, in contrast to the prevalent way where the gestures and the sequence of them have to be defined a priori and executed with exactness.

%%%%%%%%%%%%%%%%%%%%%%%%%%%%%%%%%%%%%%%%%%%%%%%%%%%%%%%%%%%%%
%%%%% References %%%%%

\bibliography{SPIE2015}   %>>>> bibliography data in report.bib
\bibliographystyle{spiebib}   %>>>> makes bibtex use spiebib.bst

\end{document}